\documentstyle[11pt]{article}
                                              
\newcommand{\SI}{\Sigma }

\newcommand{\xn}{x_{n}}

\newcommand{\e}{e^{i\int k_{0}Y}}
\newcommand{\pfe}{e^{i\ko Y + \ko .\ko [G+\SI ] + 
\ko ^V \ko ^V [\SI ]}}

\newcommand{\kim}{ k_{1}^{\mu}}                                      
\newcommand{\kom}{ k_{0}^{\mu}}                                      
\newcommand{\ki}{ k_{1}}
\newcommand{\kib}{ \bar {k}_{1}}
\newcommand{\yn}{ Y_{n}}                                             
 
\newcommand{\kn}{ k_{n}}

\newcommand{\km}{ k_{m}}

\newcommand{\kt}{ k_{2}}
\newcommand{\ktb}{\bar{k}_{2}}
\newcommand{\ko}{ k_{0}}                                             
\newcommand{\yim}{ Y_{1}^{\mu}}                                      
                                      
\newcommand{\kin}{ k_{1}^{\nu}}
\newcommand{\kon}{ k_{0}^{\nu}}                                      
\newcommand{\ktm}{ k_{2}^{\nu}}

\newcommand{\gvcp}{e^{\int dt \sum _{n\ge 0}i{\bar \kn (t)} \yn (0)+ q(t)}}
\newcommand{\dsi}{\frac{\partial}{\partial x_{1}}}

\newcommand{\dst}{\frac{\partial }{\partial x_{2}}}

\newcommand{\dsii}{\frac{\partial ^{2}}
{\partial x_{1}^{2}}}
\newcommand{\p}{\partial}

\newcommand{\eps}{ \epsilon}

\newcommand{\pin}{\mbox {$ p_{1}^{\nu}$}}

\newcommand{\kimb}{\mbox {$ \bar{k_{1}^{\mu}}$}} 
\newcommand{\kinb}{\mbox {$ \bar{k_{1}^{\nu}}$}}

\newcommand{\la}{\mbox{$ \lambda $}} 
\newcommand{\be}{\begin{equation}}
\newcommand{\br}{\begin{eqnarray}}
\newcommand{\ee}{\end{equation}} 
\newcommand{\er}{\end{eqnarray}}

\begin{document} 
\title{
\hfill\parbox{4cm}{\normalsize IMSC/2003/10/32\\
                               hep-th/0310128}\\        
\vspace{2cm}
Loop Variables with Chan-Paton Factors.}

\author{B. Sathiapalan\\ {\em Institute of Mathematical Sciences}\\
{\em Taramani                     
}\\{\em Chennai, India 600113}}                                     
\maketitle 

\begin{abstract} 
The Loop Variable method that has been developed for the U(1) bosonic open 
string is generalized to include non-Abelian gauge invariance by incorporating
``Chan-Paton'' gauge group indices. The scale transformation symmetry
$k(s) \rightarrow \la (s) k(s)$ that was responsible for gauge invariance
in the U(1) case continues to be a symmetry. In addition there is a non-Abelian
``rotation'' symmetry. Both symmetries crucially involve the massive modes.
However it is plausible that only a linear combination, 
which is the usual Yang-Mills 
transformation on massless fields, has  a smooth (world sheet) 
continuum limit. We also illustrate  how an infinite number of terms in 
the equation of motion in the
 cutoff theory add up to give a term that has a smooth continuum limit,
and thus contributes to the low energy Yang-Mills equation of motion.
\end{abstract}
\newpage      

\section{Introduction} 
A proposal for the solution to the problem of
obtaining gauge invariant equations of motion for the fields of the string
using the world sheet renormalization group [\cite{L} -\cite{Poly}],
 was described, for the free case
in \cite{BSLV}, and for the interacting case in \cite{BSGI}. A more detailed 
description was given in \cite{BSREV}.
Only the U(1) bosonic open string has been discussed so far.
Here the massless sector is described by the Maxwell action.
The logical next step is to incorporate non-Abelian symmetry. 
This is done by a variant of the Chan-Paton 
method \cite{CP}. The massless fields of this string are thus 
described, to lowest order, by the Yang-Mills action.

The Chan-Paton  method is very simple: For an M-particle amplitude,
one includes as a factor a trace of a product
of M matrices, in each cyclically inequivalent term of the amplitude. 
The matrices represents the
quantum numbers of the  external particles and are arranged in
 the same order as the corresponding vertex operators. A justification
of this method from first principles was given in \cite{MS} where dynamical 
quarks were attached to the ends of the string. This is to be contrasted 
with the heterotic string where the charges are smeared along the 
entire length of the string.

Motivated by this we implement the Chan-Paton prescription by the 
simple device of
including in the loop variable, $q(t)=q^A(t)J^A(t)$ with the properties
that $<J^A(t)>= T^A$, where $T^A$ is the matrix corresponding to the
vertex operator at $t$:
\be \label{1}
e^{\int dt \sum _{n\ge 0}i{\bar \kn (t)} \yn (0)+ q(t)}
\ee
where, as before \cite{BSGI},
\be \label{kb}
\bar \kn (t) = \sum _{ q=0}^{q=n}k_q(t)   D^n_q t^{n-q}
\ee
and 
\br
D_q^n &  = & ^{n-1}C_{q-1},\; \; n,q\ge 1 \nonumber \\
      &  = & {1\over n}, \; \; q=0 \nonumber \\
      &  = & 1 ,   \;\;           n=q=0
\er
These were defined in \cite{BSGI}.

 Thus in the case of the $U(N)$ group, $T^A$
is a Hermitian or anti-Hermitian matrix (depending on whether the mass 
level is odd or even) in the adjoint representation. The $q$'s combine
with the $\kn$'s in the loop variable and provide the space-time fields
 with the group index $A$.

Thus the following
equations and their obvious generalizations naturally suggest themselves:
\[
<\kim (t_1) q^A(t)>= A^{\mu A}\delta (t-t_1)~~~.
\]
\[
<\kim (t_1) \kin (t_2) q^A(t')q^B(t'')> = 
A^{\mu A}A^{\nu B} \delta (t'-t_1)\delta (t''-t_2)+  
A^{\nu A}A^{\mu B} \delta (t''-t_1)\delta (t'-t_2) ~~~. 
\]
\be \label{2}
<\kin (t_1) \kim (t_2) q^A(t) > = 
\delta (t_1-t_2)\delta (t-t_1)S_{1,1}^{\mu \nu A}   ~~~.
\ee

If we include $J^A(t)$ we get equations such as
\be
<\kim (t_1) q^A(t)J^A(t)>= A^{\mu A}\delta (t-t_1)<J^A(t)>= 
A^{\mu A}\delta (t-t_1)T^A ~~~.
\ee

Using the above prescription we have to repeat the steps performed
in \cite{BSGI}. This should lead to equations of motion that
correpond to strings with group theory indices. Using the same arguments
as in the U(1) case it is easy to see that when restricted to
 on shell physical states
 the S-matrix of the bosonic string supplemented with Chan-Paton
factors is reproduced. Furthermore the full equations have 
the string gauge invariance
(i.e. if we keep all the extra gauge and longitudinal states). 
Interestingly, it turns out that the non-Abelian rotation symmetry:
 $\delta A^\mu 
= [A^\mu ,\Lambda ]$ and it's extension to massive modes
is an extra symmetry! However both these symmetries involve massive modes
in a crucial way. If one wants to restrict oneself to the massless sector
- this involves taking the continuum limit on the world sheet - it is plausible
that only a linear combination survives. A study of the gauge 
transformation rules
for a (particular) massive mode does in fact support this. 
However we have not checked this
in  general. If correct this provides an interesting mechanism for
the emergence of non-Abelian Yang-Mills symmetry in string theory.  

The gauge invariant formalism requires a world sheet cutoff in a crucial
way because all the massive modes are present and being off-shell one is away
from the fixed point. The continuum limit of this theory 
is to be taken only after summing an infinite number of terms of the
 cutoff theory corresponding to contributions of an infinite number of
 irrelevant operators.
 Then
one recovers the usual S-matrix elements.

In this paper we will give a compact summary of these results, leaving
the details to a future publication. This paper uses the results of 
\cite{BSGI}. We have not included a review of previous results as that would
have made the paper too long.

This paper is organized as follows: In Section 2 we describe the loop variable
calculation of the equation of motion with Chan-Paton factors (following the
above prescription) and also show the connection with the continuum 
calculation. In Section 3 we describe the U(1) string invariance
in the presence of Chan-Paton factors and calculate the field transformation 
laws. 
In Section 4 we study the non-Abelian
rotation symmetry in terms of loop variables and also in terms of space-time 
fields.
We also discuss a mechanism mentioned above 
that might single out a linear combination 
of the two transformations in the low energy limit. 
 Section 5 contains concluding remarks.  

\section{Loop Variables with Chan-Paton Factors}

In this section we illustrate how calculations are done when the Chan-Paton
 factors are
included. Then we will show how the cubic terms arise. In the 
gauge 
invariant version we have an infinite number of terms with all powers of 
$1/\eps $
where $\eps$ is the world sheet cutoff. (By gauge invariant we mean the full 
string 
gauge invariance, not just the U(1) associated with the masselss modes.)
Gauge invariance involves massive modes also.
We will only calculate a few terms to show how gauge invariance works. To
see that the cubic  terms are there with the right coeficient as 
in the
S-matrix, one has to work with a gauge fixed version. In this case one can 
take the
continuum limit. We will show this also.

\subsection{Gauge Invariant Calculation}

Our starting point is the loop variable
\be
\gvcp
\ee
We can bring down powers of $q(t)$ as follows:
\be  \label{q}
e^{\int dt q(t)}= 1 + \int _0 ^a dt'~ q(t') + \int _0^a dt' 
\int _0^{t'}dt'' ~ q(t') q(t'') 
+ \int _0^a  dt'\int _0^{t'} dt''' \int _0^{t''} dt'''~~q(t')q(t'')q(t''') +...
\ee
The leading ``1'' will be assumed not to contribute anything. 
This is equivalent
to setting $<\kim \kin ...> =0$ when there are no $q$'s. Also the order
of the $J^A(t)$'s has to be preserved after making contractions 
of $q^A(t)$ with $\kn$'s.

{\bf Level 1}

Let us consider the level one terms:
 
\be
e^{i\ko Y + \ko .\ko [G+\SI ] + \ko ^V \ko ^V [\SI ]}
\{ \kib .\ko \dsi [G+\SI ] + i\kimb \yim \}
\int _0^a dt' q(t')
\ee
We remind the reader that $G$ refers to the (gauge covariantized)
Green function $G(z-w) = <Y(z) Y(w)>$ with coincident
arguments, viz $G(0)$. This is finite because we work with a finite cutoff. 
As
an example $G(z,w;\eps ) \approx ln~ ((z-w)^2 + \eps ^2)$ to lowest
order in $\xn$ (For more details 
see \cite{BSREV}). $\ko ^V$ in these equations has to be chosen
so that the correct RG scaling dimension is picked, which will
reproduce the continuum S-matrix term.  This is a necessary minimal
requirement. Thus in an equation for
a field of $m^2 =N$, we can set $(\ko ^V)^2 = N- (no.~of~integrations)$. 
One can make a stronger requirement:
One can require that we reproduce the RG
equation not just in the continuum limit taken after summing the
infinite series, but also for {\em finite} cutoff, term by term, before the
infinite series is summed, 
as was explained in \cite{BSREV}. 
In that case one needs to set
$\ko ^V$ equal to the powers of the cutoff $\eps$.  The correct prescription
\footnote{While the general arguments of \cite{BSREV} are correct
the precise expression for $\ko ^V$ given there, purportedly implementing the
general arguments, is not.} is
$(\ko ^V)^2 = -(no. ~of ~powers~of~{z\over \eps})-(no.~of~integrations)+
(total~(mass) ^2~of~fields)$.  
The gauge transformation
laws for space-time fields have to be defined consistent with this.
One can construct a simple algorithm that does this. 
The details will be given in a longer paper. We do not have any
 reason to prefer one option over the other.  We do however
find the latter option aesthetically
more satisfying although the former one is simpler. Perhaps further
higher order quantum consistency (loops) may pick one option.   

Varying w.r.t. $\SI$ and keeping terms proportional to $\yim$ gives 
:
\be
(\ko .\ko. \kimb - \kib .\ko \kom ) \int _0 ^a dt' q(t') =0
\ee
 Taking expectaion values:
\be
(\ko .\ko  A^{\mu A} -  \kom \ko . A ^A   )T^A  =0
\ee

This is the leading term in the Yang-Mills equation.

{\bf Level 2}

At level two massive modes make their appearance and they
also contribute to the U(1) gauge invariance. Therefore one
should not expect that the quadratic terms will have the same form
as in the Yang-Mills
equation. As explained above we will see the quadratic terms of Yang-Mills
 only when we
include all the higher levels. Nevertheless one can check that the
equations at this level are fully gauge invariant (under
the appropriate gauge transformations including massive modes).

\[
\pfe \{ \ktb . \ko \dst [G+\SI ] + \kt ^V \ko ^V \dst [\SI ] +
\]
\[
\kib .\kib \frac{1}{2} (\dsii - \dst )[G + \SI ] 
+ \ki ^V .\ki ^V \frac{1}{2} (\dsii - \dst )\SI +
\]
\[  
\underbrace 
{\frac{1}{2}(\kib . \ko \dsi [G+ \SI ] + \ki ^V \ko ^V \dsi \SI )^2}_
{1} ~+~
\]
\[ 
(\kib . \ko \dsi [G+ \SI ] + \ki ^V \ko ^V \dsi \SI ) ~ i\kimb \yim \}
\]
\be   \label{11}
(\int _0^a dt'q(t') + \int _0^a dt' \int _0 ^{t'} dt'' q(t') q(t''))
\ee

We have kept only those terms that contribute to the $\yim$ equation.

As an illustration we evaluate the contribution from the term labelled 
``1'' in (\ref{11}).
\[
\{ \kib (t_1) . \ko (t_2) ~[\dsi G]~ \kib (t_3) . \ko (t_4)  (-i\kom \yim ) \e+
\]
\[
+\kib (t_1) . \ko (t_2) ~[\dsi G]~ \ki ^V (t_3) . \ko ^V (t_4)  (-i\kom \yim )
 \e\}
\]
\be
(q^A (t') T^A + q^A(t') q^B(t'') T^A T^B)
\ee 
All the $t_i$ integrals are from 0 to $a$. The $t',t''$ integrals are
as indicated in (\ref{q}).

From (\ref{kb}) we have $\kib (t) = \ki (t) + t \ko (t)$. Let us evaluate
the leading term by replacing $\kib$ by $\ki$ and using (\ref{2}).
We get
\[
[\dsi G]\{ a S_{1,1}(\ko )^{\mu \nu A}\kom \kon (-i\ko ^ \rho )
 e^{i\ko X} T^A  ~~+
\]
\be   \label{13}
\frac{a^2}{2}(A^{\mu A}(p) A^{\nu B}(q) +
 A^{\nu A}(q)A^{\mu B}(p))(p+q)^\mu (p+q)^\nu
(-i)(p+q)^\rho e^{i(p+q)X}T^A T^B\}
\ee
Note the symmetrization in the second term. This arises because
 the contraction where $t_1=t' , t_3= t''$ gives one term and 
switching the labels
1 and 3 gives the other term.

One can similarly evaluate all the terms. Under the gauge transformation
$\kn \rightarrow \la _p k_{n-p}$ the expression in terms of loop
variables is known to be invariant before including the $(q + q^2)$
factor above \cite{BSGI}. But since this is just an overall 
multiplicative factor
it continues to be gauge invariant. What needs to be verified is that
after the contractions with $q$ are made and space-time fields are 
substituted, the gauge transformation continues to be well defined.
This proof also goes through as in the previous case because we can
recursively define gauge transformations for the massive modes in order
to make the map to space-time fields well defined. The only point one has
to worry about is whether the group transformation properties of the physical
fields at
different mass levels that we know from string theory are consistent with
these assignments.
Thus in the U(N) case (which is all we discuss in this paper)
all the mass levels are in the adjoint, but the 
odd levels are represented by Hermitian matrices whereas the even levels are
represented by anti-Hermitian matrices. This property seems to be preserved
for the physical fields at the levels that we have checked though we do 
not have a general proof. 

The level 2 term evaluated in (\ref{13}) actually vanishes because
the coefficient $\dsi G =0$. Nevertheless it illustrates the steps
involved in the calculation and all other terms can be easily evaluated
in a similar fashion.

\subsection{Gauge Fixed Calculation}

We would like to see the connection between the finite cutoff gauge
invariant calculation and the continuum limit S-matrix like calculation.
We therefore turn to the evaluation of the cubic Yang-Mills coupling.
The proper-time formalism \cite{BSPT} is convenient for this: We evaluate
\be \label{PT}
I^\mu ={d \over d ~ln~z} (z)^2 <  \p _z X^\mu (z)e^{i\ko .X(z)}~~~  
\p _w X^\nu (0)e^{ip_0 .X(0)}>\pin 
\ee
where the correlator is calculated with a nontrivial background. If $\kim$
is the coupling corresponding to the operator $\p_z X^\mu e^{i\ko .X(z)}$, 
then the
above gives ${\p S \over \p \kim}$, where $S$ is the space-time action.
A simple way to evaluate $I^\mu$  in (\ref{PT})
is to calculate $ {d \over d ~ln~z} (z)^2\kim <  \p _z X^\mu (z)
e^{i\ko .X(z)}~~~ 
 \p _w X^\nu (0)e^{ip_0 .X(0)}>\pin$
and act with ${\p \over \p \kim}$. This means that $S = \kim I^\mu $.
But this is precisely the object being calculated in the loop variable 
formalism.
Thus if we calculate an expression in the loop variable method involving three
$k$'s ($\ki (t_i)$ with i=1,2,3), and (no $Y ^\mu$'s multiplying it)
that in fact gives the (cubic term in the ) action.  

So we are led to calculate
\be
<e^{i\kim \p _z X^\mu (z) + i\ko . X (z)} 
e^{ip_1 ^\nu \p _w X ^\nu + ip_0. X(w)}
 e^{ip_1 ^\rho \p _u X ^\rho + iq_0 X(0)}>
\ee
The Chan-Paton factors coming from the 
product of three $q$'s is understood. Being an action we
will put a trace in front so that $Tr <J^A (t') J^B(t'') J^C(t''')> =
Tr (T^AT^BT^C)$. 
Let us set $p_0.q_0=k_0.q_0 =0$ for convenience.
One can do the calculation in two different ways. One way is the usual
continuum way: Thus we get a term of the form
\be
k_1.q_1 p_1.q_0 \int dw \p _z \p _u G(z,0) \p _w G(w,0) e^{k_0.p_0 
ln ~(z-w)}
\ee
The integral over $w$ produces a term of the form 
${\eps ^{k_0.p_0}\over k_0.p_0}$. If we express everything in terms of 
dimensionless $z/\eps$ then the proper time equation is just 
$\eps {d\over d\eps}{\eps ^{k_0.p_0}\over k_0.p_0}  \approx 1$ in the continuum
limit. Thus we get the cubic term in the Yang-Mills action. Note that
since $p_1.p_0 =0$ for physical states, this term has the right antisymmetry
in $k,q$, which produces a commutator of the group matrices.

Having seen the continuum limit in the gauge fixed case, 
let us expand all the Green's function
in powers of $w/\eps , z/\eps$. This is what is done in the gauge invariant
formalism. \footnote{The proper-time formalism employed a different cutoff
from what is used in the gauge invariant formalism. In the continuum
limit this should not matter. We have explicitly checked in some cases that the
arguments justifying the proper-time formalism can be used with this cutoff
also. The calculation is rather tedious and will not be reportesd here.} 
We get for the same term
\be
\ki . q_1 ~[ -{1\over \eps ^2} + {3z^2\over \eps ^4} +...]~p_1.q_0 
~[{w\over \eps ^2}~+...]
\ee
 Including the other contractions we find the leading term in $w,z$ is
\be
 k_1.q_1 (p_1.q_0 ({-w\over \eps ^4}) + p_1.k_0 {(z-w)\over \eps ^4})
\ee

If we use (\ref{kb}) we find that the linear in $w,z$ part of 
\be
{1\over \eps ^4}~\int dt_1 \int dt_2~
[~\ktb (t_1) .\ko (t_2) - {1\over 2}\kib (t_1). \kib (t_2)]
\ee
reproduces the term multiplying $k_1 .q_1$~: 
We use the notation $k (z) = k, ~~k(w) =p,
~~k(0)=q$. Note that this term is gauge invariant under the string
gauge transformations :$\ktb \rightarrow \ktb + \la _1 \kib$ and 
$\kib \rightarrow
\kib + \la _1 \ko$. It is easy to check that this is the precise 
combination that we get
in the loop variable formalism.
The higher powers of $w/\eps$ and $z/\eps$ 
are reproduced by the terms $\bar \kn . \bar \km $ The above illustrates the
connection between the gauge fixed continuum calculation and gauge invariant
cutoff calculation.

This calculation also shows that one cannot take the limit $\eps \rightarrow
 0$ in
any term because it is a singular limit. However, as the earlier
calculation shows, the infinite series has a well defined continuum limit.

\section{U(1) Gauge Transformations on Space-Time Fields}

In this section we will recursively work out the gauge transformation
 of the modes. This is a repeat of what was done in \cite{BSGI}
except for the fact that we have to be careful of the ordering
of the fields because of the matrices multiplying them.

{\bf Level 1}

At level 1 we have only the massless vectors:
\be \label{21}
\int _0^a dt_1 \int _0 ^a dt' <\kimb (t_1) q^A(t') J^A(t')> = a~A^{\mu A}T^A
\ee

The gauge transformation gives for the LHS
\be   \label{22}
\int _0^a dt \int _0^a dt_1 \int _0 ^a dt'
 <\la _1 (t) \kom (t_1)q^A(t') J^A(t')> = a~\kom \Lambda _1 ^A T^A
\ee

Comparing (\ref{21}) and (\ref{22}) we get
\be
\delta A^{\mu } = \kom \Lambda _1 
\ee
where the matrix structure is suppressed. As we explained in the introsuction
this is only the inhomogeneous part of the usual Yang-Mills gauge 
transformation.  The rotation part is missing. Nevertheless the action is
invariant because of the massive modes.

{\bf Level 2}

${\bf S_{1,1}^{\mu \nu}}$

We start with (all integrals over $t$ are understood from now on):
\be  \label{23}
<\kimb (t_1) \kinb (t_2)~ (q(t') + q(t')q(t''))>
\ee
Using (\ref{kb}) we find
\be  \label{24}
= ~a~S_{1,1}^{\mu \nu} ~+~ {a^2\over 2}~\{A^\mu , A^\nu \} ~+~ {a^2\over 2}~
\ko ^{(\mu} A^{\nu )}
\ee

The gauge transformation on the loop variable side (\ref{23}) gives:
\be  \label{25}
a~\ko ^{(\mu}\Lambda _{1,1}^{\nu )} ~+~ {a^2\over 2}\{\ko ^{(\nu}\Lambda _1, 
A^{\mu )} \}
~+~ {a^2\over 2}\{ \Lambda _1, \ko ^{(\mu} A^{\nu ) }\} ~+~
{a^2\over 2}\kom \kon \Lambda _1
\ee 
 Comparing this with the gauge transformation of (\ref{24}) gives:

\be  \label{26}
\delta S_{1,1}^{\mu \nu}~=~ \ko ^{(\mu} \Lambda _{1,1}^{\nu )} ~+~ {a\over 2}
~\{ \Lambda _1 , \ko ^{(\mu} A^{\nu )}\}
\ee
 Note that in the U(1) case this reduces to what was obtained in
\cite{BSGI}. In the anti commutator each matrix is anti-Hermitian, but the
whole term is Hermitian. This is consistent with the fact that $S_{1,1}$
is Hermitian (being at an odd mass level).

${\bf S_2^\mu}$
 
The same procedure as above gives:
\be  \label{27}
\delta S_2^\mu ~=~ \Lambda _{1,1} ^\mu ~+~{a\over 2} \{ \Lambda _1 , A^\mu \}
\ee

We see that $S_2$ is pure gauge and can be gauged away 
using $\Lambda _{1,1}^\mu$.

{\bf Level 3}

We will only present some partial results that will be used to illustrate
 arguments in the next section.

${\bf S_{2,1}^{\mu \nu}}$

We find using the above procedure:

\[
\delta S_{2,1}^{\mu \nu} ~=~ \Lambda _{1,1,1}^{\mu \nu} ~+~
 \kon \Lambda _{1,2}^\mu ~+~ {a\over 2}\{\Lambda _1 ,S_{1,1}^{\mu \nu }\}
\]
\[
~+~ {a^2\over 2} \{\Lambda _{1,1}^{\nu} , A^\mu\} ~+~ 
{a^2\over 2} \{ \kon S_2^\mu , \Lambda _1 \} ~+~ {a^3\over 8}
[\Lambda _1 , [ A^\nu , A^\mu ]]
\]
\[ ~+~ 
{a^3\over 24} \{ \Lambda _1 ,\{ A^\mu ,A^\nu \}\} 
~+~ {a^3\over 12} (A^\mu \Lambda _1 A^\nu ~+~ \mu \leftrightarrow \nu )
\]
\be   \label{28}
~+~ {a^3\over 8} [\ko ^{(\mu }A^{\nu )} , \Lambda _1 ] - {a^3 \over 24}
[\ko ^{[\mu }A^{\nu ]} , \Lambda _1 ] ~-~
 {a^3 \over 12 } [ A^\mu , \kon \Lambda _1 ]
\ee

Note that the $S_{2,1}^{(\mu \nu )}$ is pure gauge. The physical field
is the antisymmetric tensor. One can also check that if we set
$S_2^\mu =0$, then the $\Lambda _1 A A$ terms in the gauge transformation 
are symmetric in the indices $\mu , \nu$. Thus they can all be gauged to
zero. Thus the only non-trivial transformation for the physical
$S_{2,1}^{[\mu \nu ]}$ field under the $\Lambda _1$ transformation 
come from the antisymmetric parts of the last
three terms. This fact will be used in the next section.

At this level we also have $S_{1,1,1}^{\mu \nu \rho}$ which can be worked 
out similarly.

\section{Non-Abelian Rotation and Continuum Limit}

\subsection{Non-Abelian Rotation}
We saw in the previous section that when calculating the action,
 a typical term is a product of
the $k$ variables and a trace of product of $q$'s of the form
$Tr (q(t')q(t'')q(t''')...)$. This is clearly invariant under a non-Abelian
rotation of the form 
\be \label{NA}
\delta q(t) = [\tilde \Lambda , q(t)]
\ee
 for any
$\tilde \Lambda$. When we perform contractions and convert
to space-time fields this will induce non-Abelian rotations on the
space-time fields. The detailed form will depend on our choice of
$\tilde \Lambda$. The action $S$ (written in terms of space-time fields)
is also clearly invariant under this transformation. This is {\em a priori}
completely independent of the U(1) string invariance
\be   \label{U(1)}
\bar {k}(s) \rightarrow \la (s) \bar {k}(s)
\ee
 that acts on the
$\kn$ and thus we have two independent symmetries. While the gauge invariance
of the action is obvious, one has to check whether this transformation
has a well defined action on the space-time fields. Exactly as in the case
of the  U(1)-string gauge transformation, this can always be made well
defined by recursively defining the transformations of the massive fields.
Note further that this transformation leaves every monomial in $k,q$
invariant by itself, unlike the string U(1) that mixed different monomials
at a given level (but did not mix levels). Furthermore this is a {\em local}
(in space-time) rotation. Of course at the level of loop variables 
everything is ``global'' because there are no derivatives in $t$.

$\tilde \Lambda$ is arbitrary but if we want eventually to relate it to
Yang-Mills transformation we can choose it to be of the form $\approx \la  q$.
Thus let us choose 
\be
\tilde \Lambda = \int dt \int dt_0 \la _1 (t) q(t_0) f(t,t_0)
\ee
All integrals are over the range $[0,a]$ and will henceforth be suppressed.
We have chosen $\la _1$ but clearly any other $\la _n$ could have been chosen.
$f$ is an arbitrary function. One possibility is to 
choose $f = \delta (t-t_0)$. However we will let $f$ be arbitrary and try 
to fix it by requiring that the gauge transformation laws of the 
massive fields take on relatively simple forms. The choice
$f(t,t_0) = f[(t-t_0)^2]$ simplifies some of the transformations because
of the symmetry in $t,t_0$. Let us first find the transformation law
of $A$:
\[
\delta [\kim (t_1) q(t')] = \kim (t_1) \la _1(t) q^A(t_0) q^B(t') f(t-t_0)
[T^A,T^B]
\]
\be
= [A^\mu , \Lambda _1]  [f(t'-t_0) -f(0)] ~~~.
\ee
Integrals over $t_0,t'$ are understood.
Thus if we choose $f$ such that
\be   \label{A}
\int _0^a dt_0 \int _0^a dt'  [f(t-t_0) -f(0)] ~=~a
\ee
  we have
the required non-Abelian rotation
\[
\delta {NA} A^\mu =~ [ A^\mu, \Lambda _1 ,]   ~~~.
\] 

Similarly if we set 
\be  \label{B}
\int dt_0 \int dt' t_0(f(t'-t_0) -f(0)) = {a^2\over 2}
\ee
one finds the non-Abelian transformation of $S_{1,1}^{\mu \nu}$
to be as expected:
\be
\delta _{NA} S_{1,1}^{\mu \nu} ~=~[S_{1,1}^{\mu \nu},\Lambda _1] 
~+~ [A^\nu , \Lambda _{1,1}^\mu] ~+~
[A^\mu , \Lambda _{1,1}^\nu ]
\ee 
The first term is a non-Abelian rotation. The second term can be thought
of as a covariantization of  $\ko ^{(\mu} \Lambda _{1,1}^{\nu )}$.

\subsection{Continuum Limit}

We would like to suggest a possible rationale for choosing the gauge parameters
of the non-Abelian symmetry to be the same as that of the U(1) string 
invariance. We know ``experimentally'' that in string theory the low energy
sector is a non-Abelian gauge theory with the usual gauge symmetry.
In the present formalism the full theory has two independent symmetries. Both
symmetries crucially involve massive modes. This suggests the following
possibility. When we truncate the theory by keeping only the massless modes
clearly neither symmetry exists by itself but it is plausible that a linear
combination survives. This combination must then be the usual non-Abelian 
gauge symmetry. We get some support for this idea by studying
the continuum limit of the gauge transformation laws.

In order to take the continuum limit, one has to be careful about
powers of $\eps$ and $a$. The powers of $\eps$ in the loop variable
equation and the connection with the RG scaling was explained in \cite{BSREV}.
Furthermore we have seen in this paper that when we sum an infinite series the
continuum limit looks quite different. Let us give an example of this:
Consider the loop variable expression 
${1\over \eps ^2}\bar \ktm \kinb $ antisymmetrized
in $\mu ,\nu$ (The symmetric part is pure gauge). This has a piece
$S_{2,1}^{[\mu \nu ]} $ and a piece $ \approx z^2 \ko ^{[\mu} A^{\nu ]}$.
Both these terms are multiplied by $1\over \eps ^2$, corresponding to
the fact that $S_{2,1}^{[\mu \nu ]} $ is at the second mass level. 
However when we sum an infinite series arising from terms of the form
${1\over \eps ^{n+m}} \bar \kn . \bar \km $, 
 it is possible that the
second term sums to a completely different value. As an illustration
suppose the first term is just $1\over \eps ^2$ but the second one is the first
term in the following series:
\[ 
{z^2\over z^2 + \eps ^2} = {z^2\over \eps ^2} - {z^4 \over \eps ^4} +...
\]
If one acts with $\eps {d\over d\eps}$ the leading term gives -2. 
However after summing the series the same operation gives zero, in the limit
$\eps \rightarrow 0$. Thus typically,
terms in a loop variable expression 
involving different powers of $z$, belong to different infinite series, 
which have different behaviours in the 
limit $\eps \rightarrow 0$ after being summed.

Keeping this in mind, if one looks 
for terms involving $z$ in the transformation laws of physical massive fields
(i.e. that cannot be gauged away), we find that they occur for the first time 
in the case of the level-3 field, $S_{2,1}^{[\mu \nu ]}$.  These 
 are the commutator terms in the last line of (\ref{28}). Thus there is a 
possibility
that this term is the first in a series such that the limit 
$\eps \rightarrow 0$ for this series is very different
from $1\over \eps ^2$.
     (The rest of the terms in (\ref{28}) 
do not involve massless fields or can be gauged away as explained
in the paragraph below (\ref{28}).)  Thus it is possible that the gauge 
transformations, in the limit $\eps \rightarrow 0$ produce two kinds of terms:
one goes as $1\over \eps ^2$ and the other $\rightarrow 1$ .  That would mean
the massive modes remain coupled by its gauge transformation to the massless
sector.   
However if we now include the non-Abelian rotation
we get the following extra terms:
\[
[\kon A^\mu , \Lambda _1] (t_0^2f(t'-t_0) ~-~ t'^2 f(0)) ~+~
[A^\mu , \kon \Lambda _1 ]t't_0(f(t'-t_0) -f(0)) ~+~
\]
\be
[\kom A^\nu , \Lambda _1] {1\over 2}(t_0^2f(t'-t_0) ~-~ t'^2 f(0))~+~
[A^\nu , \kom \Lambda _1 ]{1\over 2}(t_0^2f(t'-t_0) ~-~ t'^2 f(0))
\ee
By choosing 
\br  \label{C}
(t_0^2f(t'-t_0) ~-~ t'^2 f(0)) ~=~{a^3\over 6}\nonumber\\
({t't_0\over 2} - {t'^2 \over 4})(f(t'-t_0) -f(0) ) ~=~ {a^3\over 6}
\er
one finds that the offending commutator terms can be cancelled. Since $f$ is a 
completely arbitrary function we should be able to satisfy the constraints
(\ref{A}),(\ref{B}) and (\ref{C})
that have been imposed on it thus far.

Thus we have the
possibility that with a non-zero $f$ (and hence a gauge field having the
usual non-Abelian transformation) there is a well defined limit when the theory
is truncated to massless fields. 
     We hasten to add that the above discussion of a mechanism that could
single out the linear combination of the two symmetries as the symmetry of the
low energy theory with just massless fields, is very tentative.
 Only a detailed study of the continuum
limit will show whether this suggestion is correct.  

\section{Conclusions}

In this paper we have described a proposal for incorporating Chan-Paton factors
in the loop variable formalism. As in the U(1) case the construction
is guaranteed to reproduce on-shell S-matrix elements of the corresponding
string theory. The action with all the massive modes present has in addition
to the U(1) string invariance, non-Abelian ``rotational''
 invariance. This includes
the usual homogeneous non-Abelian transformations 
 of the space-time fields as a special
case. We have also discussed a mechanism which might explain why
a linear combination of the U(1) string transformations and the non-Abelian
transformations- which is the usual non-Abelian
gauge transformation - is what survives at low energies as a symmetry 
of an effective action that describes just the massless modes. 

To make the argument rigorous one would need to study the continuum limit
for all the modes. Clearly some field redefinitions are needed to make this 
low energy limit look more natural.
We also need to study other gauge groups in more detail. Finally, and most
important, this construction should be extended to closed strings. This will
involve replacing $J^A(t)$ by an anti-holomorphic sector.
We hope to report on these issues in the future. 

\vspace{5mm}
{\bf Acknowledgements}

Part of this work was done while the author was visiting the AS
International Centre for Theoretical Physics, Trieste, Italy. 
We would like to thank the 
members of the High Energy Group for their hospitality.

\end{document}